\documentclass[journal]{IEEEtran}

\usepackage{graphicx}
\usepackage{color}
\usepackage{placeins}
\usepackage{float}
\usepackage{tabularx,colortbl}
\usepackage{amssymb}
\usepackage{amsthm}
\usepackage{algorithm}  
\usepackage{algorithmic}
\usepackage{cite}
\usepackage{amsmath}
\usepackage{makecell}
\usepackage{subfigure}
\usepackage{caption2}
\usepackage{enumitem}

\theoremstyle{plain}

\theoremstyle{plain}

\IEEEoverridecommandlockouts

\begin{document}
	\title{Rydberg Atomic Receivers for Wireless Communications: Fundamentals, Potential, Applications, and Challenges}
	\author{Yin Zhang, Jiayi Zhang,~\IEEEmembership{Senior Member,~IEEE}, Bokai Xu, Yuanbin Chen, Zhilong Liu, Jiakang Zheng, \\Enyu Shi, Ziheng Liu, Tierui Gong,~\IEEEmembership{Member,~IEEE}, Wei E. I. Sha,~\IEEEmembership{Senior Member,~IEEE}, \\Chau Yuen,~\IEEEmembership{Fellow,~IEEE}, Shi Jin,~\IEEEmembership{Fellow,~IEEE}, and Bo Ai,~\IEEEmembership{Fellow,~IEEE}
	\thanks{Y. Zhang, J. Zhang, B. Xu, Z. Liu J. Zheng, E. Shi, Z. Liu and B. Ai are with the School of Electronics and Information Engineering, Beijing Jiaotong University, Beijing, 100044; W. E. I. Sha is with the College of Information Science and Electronic Engineering, Zhejiang University, Hangzhou, 310027; Y. Chen, T. Gong and C. Yuen are with School of Electrical and Electronics Engineering, Nanyang Technological University, Singapore, 639798; S. Jin is with the National Mobile Communications Research Laboratory, Southeast University, Nanjing, 210096.}
	}
	\maketitle
	\begin{abstract}
	Rydberg atomic receivers (RARs) leverage the quantum coherence of highly excited atoms to overcome the intrinsic physical limitations of conventional radio frequency receivers (RFRs), particularly in sensitivity, and bandwidth. This innovative technology represents a paradigm shift in wireless communication systems. This paper systematically explains the fundamental sensing mechanisms of RARs, contrasts their differences from RFRs in working principles and architectures. We explore their advantages in emerging wireless communication scenarios, such as integrated sensing and communications, quantum Rydberg radar, and quantum space communications. Practical challenges, such as limited instantaneous bandwidth and nonlinear distortion, are identified. To address these issues, mitigation strategies and future research directions are also outlined, supporting the advancement of RAR-aided wireless systems.
	\end{abstract}
	

	\section{Introduction}
	For modern wireless communication systems, industry and academic researchers are searching for new revolutionary wireless network technologies to meet the growing demand for higher transmission rates, greater reliability, and lower power consumption. However, conventional radio frequency receivers (RFRs) which rely on classical  classic electromagnetic induction and semiconductor physics principles, encountering fundamental roadblocks in sensitivity, noise characteristics, usable bandwidth, and antenna coupling. In fact, Chu's gain-bandwidth constraint and Hannan's efficiency limit \cite{10918774} mathematically bound the tradeoffs among these key parameters, so that no amount of incremental circuit or materials engineering can simultaneously push all figures of merit forward. Consequently, meeting future requirements will demand truly revolutionary receiver architectures and network technologies, rather than further tweaks to today’s RFR designs.
	
\begin{figure*}[!t]
		\centering
		\includegraphics[width=6in]{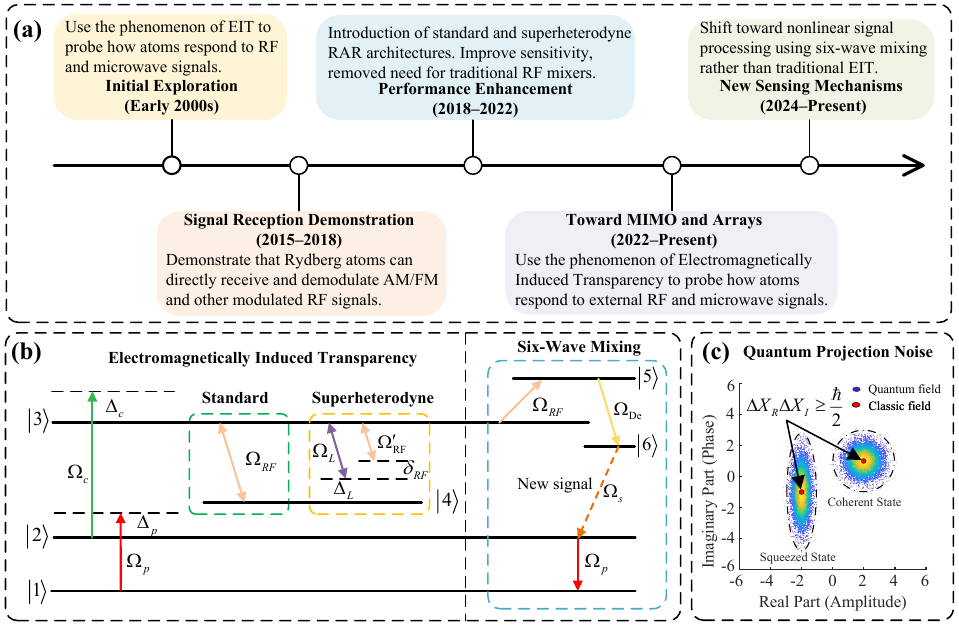}
		\caption{(a) The evolution of Rydberg atomic receiver. (b) The Energy level diagrams of different sensing mechanisms. (c) Comparison between classic and quantum electromagnetic fields.}
		\label{fig:Sensing_Mechanism}
	\end{figure*}

	In order to achieve more precise sensing and improve the received signal-to-noise ratio (SNR), Rydberg atomic receivers (RARs) can be innovatively introduced into the next-generation wireless communication architectures. As illustrated by the evolution of RAR over recent decades, shown in Fig. \ref{fig:Sensing_Mechanism}(a), these receivers offer a promising path forward. Compared to RFRs, RARs demonstrate breakthrough capabilities in sensitivity (approaching the standard quantum limit (SQL)), operating bandwidth (spanning $\rm{DC}$ to $\rm{TH}z$ frequencies), and electromagnetic field detection range (from $\mu\rm{V}/m$ to $k\rm{V}/m$) \cite{Legaie_2024, santamariabotello2022comparisonnoisetemperaturerydbergatom,kim2024quantummusicmultiplesignalclassification,10972179,10238372,Holloway2022}. Such a quantum-enabled performance leap derives from exploiting atomic-scale interaction mechanisms that entirely bypass the classical induction process, unlocking a fundamentally new paradigm for wireless communications.

	RARs based on this quantum architecture has the following key advantages compared to conventional RFRs.
		
	\textbf{$\bullet$ Higher Sensitivity:} Rydberg atoms possess an exceptionally high polarizability that scales as \(n^7\) (with \(n\) being the principal quantum number) \cite{10238372}, allowing them to acquire electric dipole moments on the order of a Debye under alternating electric fields. This remarkable dipolar response endows RAR systems with significantly enhanced signal reception capabilities and sensitivity \cite{Holloway2022}.

	
	\textbf{$\bullet$ Broadband Response:} RARs support broadband signal detection by utilizing both resonant and off-resonant atomic transitions within the Rydberg states. Through optical tuning of the atomic levels, RARs can operate across an ultra-wide spectral range. In contrast, traditional RFRs are constrained to narrow frequency bands due to the limited bandwidth of electronic components such as amplifiers and filters \cite{Deb_2018}.
	
	\textbf{$\bullet$ Non-Invasive Field Detection:} By relying exclusively on optical interrogation of atomic energy levels within a vapor cell, RARs enable entirely passive, contact-free detection of ambient electromagnetic fields. This capability is especially valuable for applications that demand minimal perturbation of the measured field or operation within electromagnetically sensitive environments.
	
	Building upon these foundational advantages, this paper systematically investigates the potential of Rydberg atomic receivers in wireless communication systems. We first introduce the sensing mechanisms and structural characteristics of RARs, highlighting their distinctions from conventional receivers. We then explore representative application scenarios, including integrated sensing and communications, quantum Rydberg radar, and quantum space communications. Finally, we highlight some practical challenges and offer insights into potential solutions and research directions for the effective deployment of RARs.
	
	\section{Sensing Mechanism of Rydberg Atomic Receivers}
	To provide a comprehensive understanding of RARs, this section commences by examining two sensing mechanisms of Rydberg atoms. These mechanisms clarify the interactions between the Rydberg atoms and radio frequency (RF) fields, which facilitates an accurate readout for RF field measurements by RARs. 
	
	\subsection{Electromagnetically Induced Transparency}	
	
	RAQRs can leverage the phenomenon of electromagnetically induced transparency (EIT) to enable highly sensitive detection of the weak signals. The core mechanism involves constructing a multi-level atomic system in which the Rydberg state serves as the final excitation level. Specifically, a probe laser drives the atomic transition from the ground state $|1\rangle$ to an intermediate state $|2\rangle$, while a counter propagating coupling laser promotes the subsequent transition from $|2\rangle$ to a highly excited Rydberg state $|3\rangle$, as shown in Fig. \ref{fig:Sensing_Mechanism}(b).
	Under two-photon resonance conditions, quantum interference between the two excitation pathways leads to a dramatic suppression of absorption at the probe frequency, rendering the vapor cell transparent to the probe laser. 
	By scanning the coupling laser and monitoring the resultant probe absorption, the EIT signal effectively spectroscopically probes the energy of the Rydberg state $|3\rangle$, as the probe absorption decreases when the coupling laser is aligned with the transition energy \cite{chen2025harnessingrydbergatomicreceivers}.
	
	Due to the highly excited nature of Rydberg states, the system exhibits extreme sensitivity to external microwave fields. Even very weak electromagnetic signals can perturb the coherence of the system by coupling transitions between nearby Rydberg levels, thereby modifying the EIT spectral profile. Thus, EIT not only serves as a fundamental quantum-optical phenomenon but also underpins the core operating principle of Rydberg-based atomic receivers. When an RF field induces a resonant coupling between $|3\rangle$ and $|4\rangle$, the resultant autler townes (AT) effect manifests itself as the energy level splitting defined by the Rabi frequency, as depicted in Fig. \ref{fig:Sensing_Mechanism}(b), which is given by $\Omega_{\text{RF}}(t) = \mu_{\text{RF}} E(t) / \hbar $, where \(\hbar\) is the reduced Planck constant, \(\mu_{\text{RF}}\) denotes the electric dipole moment, \(E(t)\) is the time-dependent electric field, and \(\delta\) represents the detuning between the RF frequency \(f\) and the atomic transition frequency \(f_{|3\rangle \to |4\rangle}\) \cite{10238372}. This expression highlights how the electric field \(E(t)\) is effectively amplified through its interaction with the dipole moment \(\mu_{\text{RF}}\).
	
	\begin{figure*}[!t]
		\centering
		\includegraphics[width=5in]{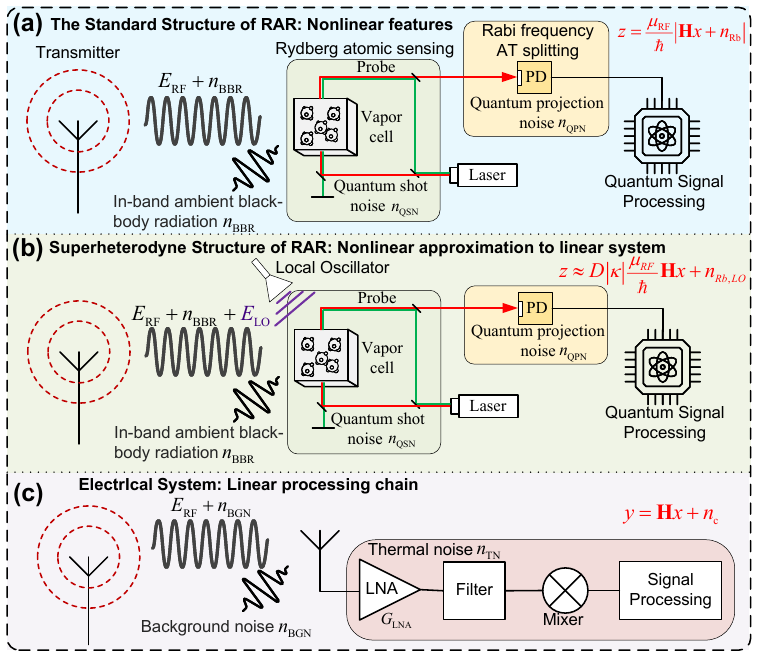}
		\caption{The architecture comparison between quantum systems and electrical systems. (a) The Standard Structure of Rydberg Atomic Receiver. (b) Superheterodyne Structure of Rydberg Atomic Receiver. (c) Traditional radio frequency receiver.}
		\label{fig:RbMIMO}
	\end{figure*}
	
	\subsection{Six-Wave Mixing}
	Six-wave mixing represents a nonlinear optical process that enables continuous-wave microwave-to-optical conversion at room temperature without requiring cryogenic conditions or tightly controlled atomic trapping. As shown in Fig. \ref{fig:Sensing_Mechanism}(b), in this scheme, Rydberg atoms in a vapor cell mediate a six-field interaction, involving three optical fields including probe, coupling, and decoupling lasers, and an incident RF field. These fields collectively drive a closed-loop excitation among atomic states, culminating in the emission of a signal photon at a distinct optical frequency \cite{Bor_wka_2023}. The six-wave mixing process hinges on energy and momentum conservation, with wavevectors and polarizations carefully phase-matched to ensure coherent buildup of the emitted optical field. Compared to EIT-based sensing, six-wave mixing provides a broader bandwidth (up to tens of MHz) and operates continuously thanks to the constant influx of thermal ground-state atoms in the vapor cell. The signal strength scales with the microwave electric field amplitude.
	
	
	A key feature of this approach is its robustness against noise. Since the converted signal emerges at a spectrally distinct wavelength, optical filtering effectively isolates it from background fields. Compared with the EIT-AT-based sensing mechanism, the six-wave mixing approach exhibits several notable distinctions. First, while EIT-AT relies on monitoring the transmission spectrum of a probe laser under the influence of microwave-induced Rydberg state transitions, six-wave mixing directly generates a new optical field through nonlinear wave mixing, enabling microwave-to-optical frequency conversion. This key difference means that six-wave mixing does not depend on detecting absorption changes but instead produces an upconverted signal that can be isolated and measured via photon counting. In terms of signal isolation and robustness, six-wave mixing offers superior spectral separation, as the converted signal lies at a distinct wavelength, allowing effective filtering of background noise and stray laser light. By contrast, EIT-AT systems often face limitations in separating the probe from coupling-induced artifacts. six-wave mixing supports continuous-wave operation in warm atomic vapors, allowing uninterrupted monitoring of weak or asynchronous microwave fields. In contrast, EIT-AT setups are often constrained by atomic coherence times and may require time-gated detection schemes. These attributes make six-wave mixing particularly suitable for broadband microwave photonic conversion and quantum-level field sensing in ambient conditions, expanding the application space beyond EIT-ATs. However, the implementation of six-wave mixing generally requires more complex optical configurations, stringent phase-matching conditions, and additional laser fields, resulting in higher system complexity and deployment costs compared to the relatively simpler EIT-ATs based receivers.
		
	\section{The Architectures of Rydberg Atomic Receiver}
	This section presents the information transfer mechanism employed in RARs, in comparison with conventional RFRs. Particular emphasis is placed on the standard RAR configuration and superheterodyne configuration, with a detailed discussion of their architectural differences and respective advantages.
	\subsection{The Standard Structure of Rydberg Atomic Receiver}
\begin{figure}[!t]
		\centering
		\includegraphics[width=3.5in]{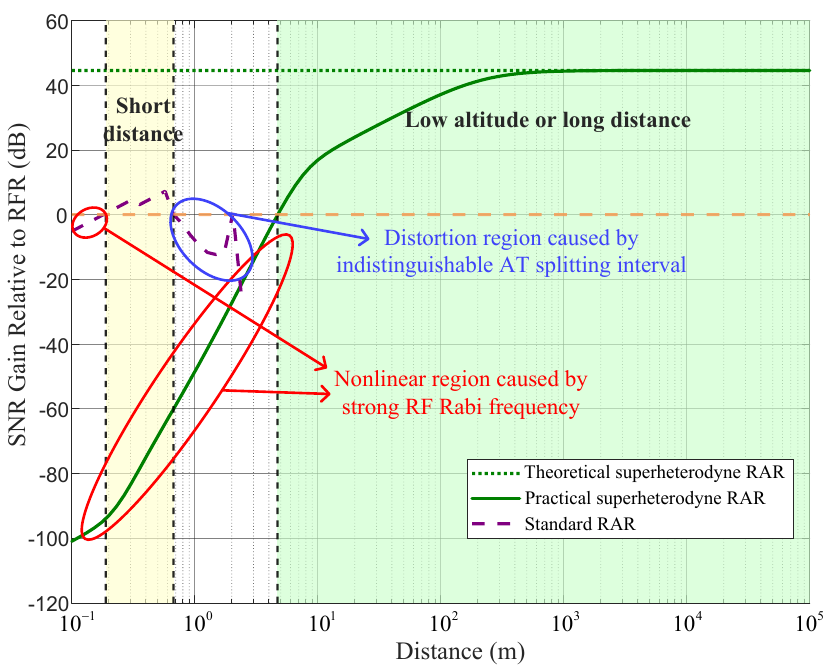}
		\caption{SNR gain versus distance of Rydberg atomic receiver compared to traditional receivers}
		\label{fig:SNR_gain_vs_Dist}
	\end{figure}

	Fig. \ref{fig:RbMIMO}(a) shows the standard configuration of RAR, also known as the local oscillator (LO)-free deployment, which leverages EIT AT effect to detect RF signals. Unlike conventional RFRs, which absorb RF energy and convert it into conduction currents for amplification and processing, the standard RAR does not absorb the RF field. Instead, it coherently modifies the atomic energy levels through RF signals, and applies amplitude modulation on the probe laser, which is subsequently measured via optical methods \cite{Holloway2022, Legaie_2024}.
	Compared to the structure of RFR shown in Fig. \ref{fig:RbMIMO}(c), this optical readout approach eliminates the need for bulky electronic front-end components, simplifying the receiver architecture and significantly reducing power consumption \cite{Song2019}. However, the standard RAR is limited to amplitude detection and cannot resolve the phase information of RF signals, restricting its applicability in phase-modulated communication systems. Moreover, as shown in Fig. \ref{fig:SNR_gain_vs_Dist}, under extremely weak signal conditions, the ATs becomes indistinct, pushing the system into its distortion region with degraded performance \cite{chen2025harnessingrydbergatomicreceivers}.
	
%
	
	\subsection{Superheterodyne Structure of Rydberg Atomic Receiver}
	The superheterodyne RAR configuration, also termed the LO-dressed configuration. As shown in Fig. \ref{fig:RbMIMO}(b), it builds on the standard RAR by incorporating a LO to enable both amplitude and phase detection, mirroring the superheterodyne principle of traditional RFRs but employing distinct quantum-based mechanisms. 
	Architecturally, the superheterodyne  configuration introduces an additional RF field serving as the LO. Compared to traditional superheterodyne RFRs, which rely on antennas, LNAs, and mixers, the RAR integrates these functionalities into a compact atomic system. The absence of energy absorption and the use of optical detection eliminate the need for front-end amplification, filtering, and mixing, thereby reducing size and complexity. Compared to the standard RARs, the superheterodyne configuration introduces the second RF field that requires precise control to maintain resonance with the Rydberg transition \cite{chen2025harnessingrydbergatomicreceivers}.
		
	The information transmission mechanism in the superheterodyne RAR still depends on the interplay of EIT and ATs. This configuration acts as an atomic mixer, converting the RF signal to an intermediate frequency by beating the incoming signal with the LO. This process parallels the frequency conversion in traditional superheterodyne RFRs, which use electronic mixers for down-conversion. RAR can obtain the frequency and phase differences between the RF signal and LO via optically tracking \cite{gong2024rydbergatomicquantumreceivers}.	
	However, the ability of superheterodyne RARs to accurately acquire RF information relies on the assumption that the power of LO is stronger than the signal. As shown in Fig. \ref{fig:SNR_gain_vs_Dist}, in the presence of strong RF signals at short distances, the system is prone to entering its nonlinear region, leading to signal distortion and performance degradation. Traditional superheterodyne RFRs typically mitigate strong signal interference through gain control, but such designs require complex circuitry to achieve sensitivity comparable to that of RARs. In contrast, by leveraging the complementary strengths of short-range standard RAR and long-range superheterodyne RAR, RARs offer a versatile solution for next-generation wireless systems, surpassing the limitations of traditional RF receivers in terms of sensitivity, flexibility, and compactness.
			
	\subsection{Noise Model of Rydberg Atomic Receivers}	
	The noise in RARs as random disturbances affecting quantum states. The noise intensity, denoted by $\lambda$, determines the probability that the quantum state degrades into a maximally mixed state \cite{gong2024rydbergatomicquantumreceivers}. This explanation is based on an abstract perspective from information theory, which neglects the physical significance of noise. In this subsection, we will summarize the sources of noise in RARs.
	
	\textit{1) External Noise Sources:} The external noise mainly consists of black-body radiation and quantum fluctuations \cite{santamariabotello2022comparisonnoisetemperaturerydbergatom}. 
	In-band ambient black-body radiation arises from the thermal agitation of particles in the surrounding environment. The volumetric energy density of the thermally generated electromagnetic field at any point \( r_0 \), observed within a narrow bandwidth \( \Delta f \), is given by \( \langle W_{\text{b}} \rangle = \frac{8\pi h f^3 \Delta f}{c^3} n_{0}(f, T) \), where \( h \) is Planck constant, \( f \) is the signal frequency, and $n_{0}(f,T)$ is the Bose-Einstein distribution \cite{santamariabotello2022comparisonnoisetemperaturerydbergatom}. For RARs, this in-band radiation is an unavoidable factor.
	In a superheterodyne setup, the actual thermal energy is twice that of \( \langle W_{\text{b}} \rangle\) because of the downconverted noise. In a standard setup, the actual thermal energy is half of \( \langle W_{\text{b}} \rangle\) because only one quadrature of the electromagnetic field is observed and both are equally thermally populated.
%

	\textit{2) Inherent Noise of the RARs:} Unlike RFRs, the RARs avoids thermal noise and nonlinear distortions from components like low-noise amplifiers (LNA) and mixers. Instead, their sensitivity is limited by two types of quantum-originated noise: quantum shot noise (QSN) and Quantum projection
	noise (QPN) \cite{santamariabotello2022comparisonnoisetemperaturerydbergatom}. 
	

	\begin{figure*}[!t]
		\centering
		\includegraphics[width=6.3in]{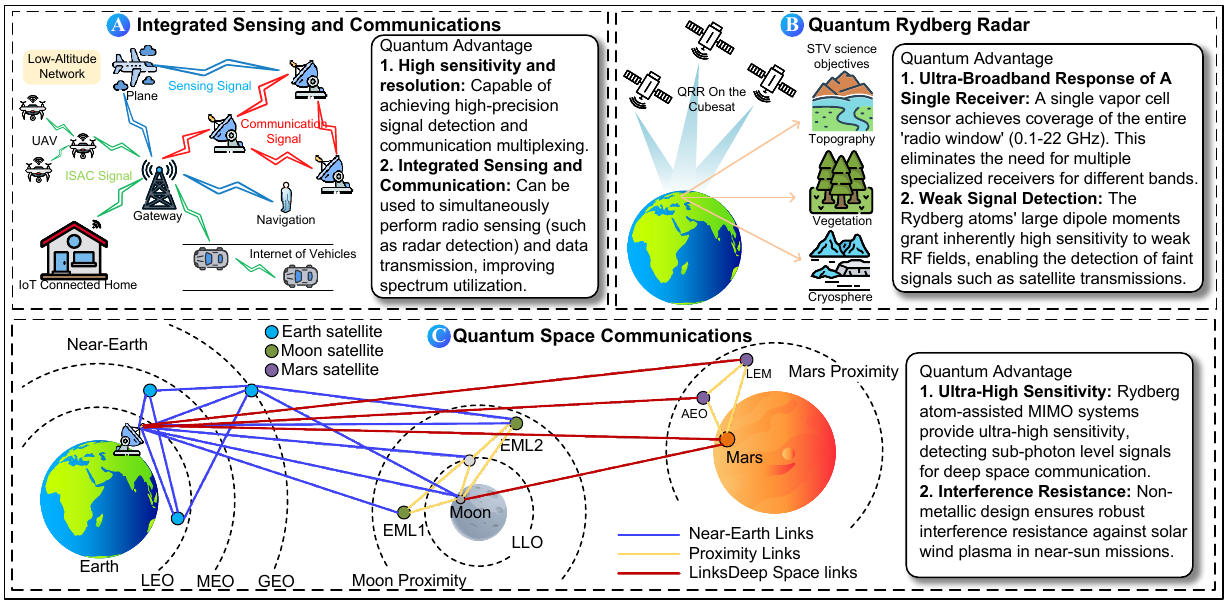}
		\caption{Potential applications of Rydberg atom-aided MIMO systems for wireless communications.}
		\label{fig:Application}
	\end{figure*}
	
	QSN arises during the optical detection process and stems from the discrete nature of photons. When probing the atomic response using laser beams, the number of detected photons fluctuates due to Poissonian statistics, leading to intensity noise in the measurement. This type of noise becomes dominant when photon numbers are low or detection time is short. The QSN can be suppressed by employing techniques such as optical frequency comb (OFC) stabilization, which can reduces laser amplitude and phase noise. 
	
	QPN originates from the quantum measurement of atomic states. When a large ensemble of atoms is prepared in a superposition state and measured collectively, the projection of each atom onto a Rydberg state introduces statistical fluctuations. This leads to uncertainty in estimating population distributions, as illustrated in Fig. \ref{fig:Sensing_Mechanism}(c). Both noise sources are several orders of magnitude below thermal noise in conventional systems, enabling RARs to achieve exceptional sensitivity \cite{Holloway2022}. Their impact can be reduced by increasing the number of atoms, extending the coherence time, and improving readout fidelity.

	\section{Application Scenarios of Rydberg Atomic Receivers}
	RARs enable novel capabilities in wireless systems through quantum-enhanced sensing, ultra-broadband reception, and high sensitivity. This section highlights their potential in critical application domains, including integrated sensing and communication (ISAC), quantum-enhanced radar, and deep-space communication. Each use case demonstrates how RARs address conventional limitations, as illustrated in Fig. \ref{fig:Application}.
		
	\subsection{Integrated Sensing and Communications}
	RARs demonstrate exceptional potential in integrated sensing and communications scenarios. Their high sensitivity and quantum resonance properties enable accurate signal detection in weak signal environments \cite{Holloway2022}. Compared to RFRs, which face challenges related to signal attenuation and interference when performing simultaneous communications and sensing tasks, RARs efficiently perform communications and sensing on a single platform. For instance, in autonomous driving or smart cities, they support data transmission and precise target detection without additional hardware, reducing costs and complexity. 
	Additionally, quantum-enhanced processing further strengthens anti-interference performance, making them ideal for vehicular networks where they enable simultaneous communications and radar sensing \cite{10972179,xu2025channelestimationrydbergatomic}.
	
	\subsection{Quantum Rydberg Radar}
	Quantum Rydberg Radar (QRR) extends the application landscape of RARs by enabling them to function as core detection elements within next-generation radar systems. Rather than replacing full radar architectures, RARs provide a quantum-enhanced front-end that significantly expands radar performance, especially in missions requiring wide frequency coverage, compact size, and high sensing fidelity.
	﻿
	RARs are inherently compact and lightweight, consisting primarily of vapor cells and laser optics without the need for large antenna structures or cryogenic systems. This compactness is particularly valuable for size, weight, and power constrained platforms such as high-altitude balloons, CubeSats, or UAVs. Additionally, RARs exhibit ultra-broadband detection capability, enabling continuous reception from sub-GHz frequencies up to hundreds of GHz within a single sensing module. This removes the need for traditional band-specific hardware chains and supports frequency-agile radar operations, where sensing bands can be reconfigured dynamically in response to mission demands or environmental changes.
	
	\subsection{Quantum Space Communications}
	In quantum space communications, RARs have demonstrated breakthrough potential. Classic systems face challenges such as signal attenuation, latency sensitivity, and security in long-distance transmission. With sub-photon sensitivity, RARs capture weak signals in deep space, such as in Earth-Moon laser links \cite{Holloway2022}. Unlike traditional antennas limited by diffraction and noise, Rydberg vapor cells achieve sub-wavelength precision \cite{Liu_2022}, enabling accurate beam alignment for compact spaceborne terminals. In addition, non-metallic design also avoids solar wind plasma interference, ensuring stable near-sun communications \cite{Deb_2018}.
	
	\begin{figure}[!t]
		\centering
		\includegraphics[width=3.5in]{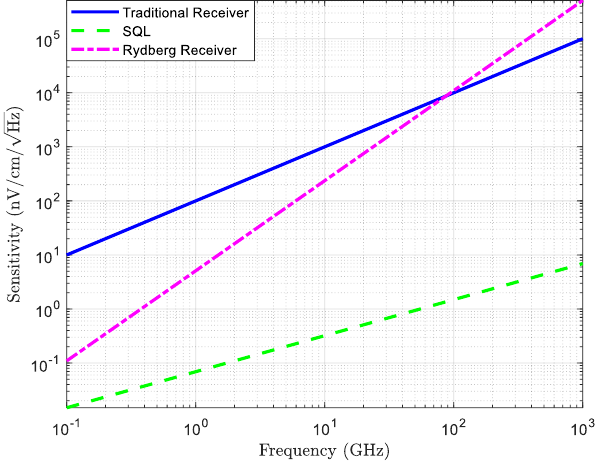}
		\caption{Comparison of sensitivity between traditional receiver and Rydberg atomic receiver.}
		\label{fig:Sensitivity}
	\end{figure}
	
	\section{Challenges}
	\subsection{Instantaneous Bandwidth Widening}
	The instantaneous bandwidth of RARs is typically limited to around 10 MHz, primarily due to constraints imposed by the EIT linewidth, nonlinear signal transduction, and stabilization delays associated with dynamic laser tuning \cite{cui2024rydbergatomicreceiverfrontier}. This narrow instantaneous bandwidth significantly restricts the achievable data rates of RARs in communication applications.
	
	To address this limitation, several enhancement strategies are proposed. One approach is to employ multiple probe lasers by using spatiotemporal multiplexing to enhance the instantaneous bandwidth of the RARs during the EIT process. Furthermore, nonlinear optical techniques such as six-wave mixing could be implemented to suppress noise, enhance high-frequency response, and broaden effective bandwidth. Alternatively, parameter optimization methods such as minimizing the laser waist and enhancing the coupling laser power can be used to broaden the EIT bandwidth, thereby improving instantaneous bandwidth and response speed of RARs.
	
	\subsection{Sensitivity Further Improved}
	As shown in Fig. \ref{fig:Sensitivity}, RARs have already achieved considerable sensitivity, but there remains a gap to the SQL. The SQL, constrained by QSN and QPN, sets a lower bound on the smallest detectable field. In practice, RAR sensitivity is further degraded by atomic density and technical noise: excessively high atomic density leads to atom–atom and wall collisions that broaden spectral lines and shorten coherence time, while overly low density raises SQL. Likewise, thermal motion (Doppler and transit-time broadening) and laser technical noise (finite linewidth, power fluctuations) impose additional limits on coherence and detection precision.
	
	To push toward the SQL, one must both suppress these noise sources and harness quantum resources.  For example, using laser-cooled or buffer-gas vapor cells and carefully optimized atom density can balance collisional broadening against projection noise, and employing ultrastable narrow-linewidth lasers with power stabilization can nearly eliminate probe noise. Additional techniques like advanced signal-processing (e.g. machine-learning algorithms that filter out environmental perturbations) have also been proposed to drive sensitivity toward the theoretical limit.
	\subsection{Handling of Distortion Regions}
	As discussed in Section III, when the incident RF power becomes too high, the system enters a nonlinear regime characterized by saturation effects, power broadening, and multi-level coupling. In this regime, the Rabi frequency associated with the RF field no longer scales linearly with the EIT peak splitting, causing ambiguity in amplitude measurements and degrading the signal fidelity. Furthermore, strong fields may induce undesirable multi-photon transitions or mixing between neighboring Rydberg states, leading to spectral distortion and cross-talk. On the other hand, extremely weak signals may push the system below the resolvable threshold of AT splitting, where the peak interval becomes indistinguishable from noise, especially in the standard configuration. These nonlinear effects constrain the RAR’s dynamic range.
	
	To address this issue, a number of strategies can be pursued. Dynamic gain control, similar to automatic gain control in conventional RF systems, can be implemented to maintain operation within the linear region by adjusting laser power or field amplitude in real time.
	
	\subsection{Rydberg Atom-aided MIMO}
	﻿Rydberg-atom-aided MIMO systems can be realized via two primary approaches: deploying multiple vapor cells as spatially separated receiver elements, or utilizing a single vapor cell with multiplexed probe beams to emulate parallel channels. While both architectures aim to exploit spatial diversity and multiplexing gains, they face distinct implementation challenges. In the multi-cell configuration, each vapor cell operates with its own set of probe and coupling lasers, enabling true spatially distributed reception. However, this introduces substantial complexity in terms of optical alignment, laser frequency stabilization, and calibration across channels. Any mismatch in cell conditions, such as temperature, atomic density, or stray electromagnetic fields, can degrade channel orthogonality and compromise array performance. Moreover, maintaining uniform sensitivity and phase coherence across all cells becomes increasingly difficult as the number of channels grows. On the other hand, the single-cell multiplexed design offers a more compact and hardware-efficient alternative by injecting multiple probe beams into distinct spatial regions within a shared vapor volume. This reduces physical complexity but introduces challenges in channel separation due to atomic motion, state diffusion, and inter-beam coupling. The spatial resolution of virtual channels may degrade over time, limiting scalability and robustness.
	
	\section{Conclusions}
	This paper provides a comprehensive study of RARs, including their sensing principles, architectural designs, application prospects, and associated technical challenges. By harnessing the quantum coherence of Rydberg atoms, RARs enable enhanced sensitivity, broad frequency coverage, and compact receiver designs that go beyond the limitations of classical RF front-end. Application scenarios such as ISAC, multi-band communication, and deep-space links highlight the transformative potential of this technology. The limitations of RARs, such as a relatively narrow instantaneous bandwidth, insufficient sensitivity approaching the quantum limit, and nonlinear response under strong or weak fields, have also been discussed. Although Rydberg atom-aided MIMO systems are still in the early stages of exploration, they represent a promising direction for future research. Future work should focus on enhancing signal processing robustness, optimizing quantum coherence, and improving deployment scalability.

	\bibliographystyle{IEEEtran}
	\bibliography{IEEEabrv,Ref}
\end{document}